# CLOUD COMPUTING ADOPTION: OPPORTUNITIES AND CHALLENGES FOR SMALL, MEDIUM AND MICRO ENTERPRISES IN SOUTH AFRICA

Simphiwe S. Sithole, University of Witwatersrand, South Africa, sssithole@hotmail.com

Ephias Ruhode, Cape Peninsula University of Technology, South Africa, RuhodeE@cput.ac.za

**Abstract:** The purpose of the paper is to determine the opportunities and challenges that lead to cloud computing adoption by SMMEs in South Africa by looking at the factors that influence adoption. The TOE framework is used to contextualize the factors that influence cloud computing adoption and evaluate the opportunities and challenges that are presented by cloud computing to SMMEs in South Africa. An online survey questionnaire was used to collect data from leaders of SMMEs from all geographical regions and business industries in South Africa. A quantitative research approach was adopted to investigate the objectives, and descriptive analysis was used to evaluate the relationships and present the results. The findings of the study show that relative advantage is an important factor in the consideration of cloud computing adoption by SMMEs, while government and regulatory support is perceived as a barrier. Top management support, which has been previously found by other studies to be a significant factor has been found to be insignificant in this study. The study has revealed that cloud computing presents opportunities to SMMEs and improves their competitiveness.

## 1. INTRODUCTION

Small, medium and micro enterprises (SMMEs) are critical to the success of the economies of many countries around the world. They contribute to the creation of employment, the growth of the gross domestic product (GDP), the development of professional skills and the reduction of poverty. However, in South Africa SMMEs face many challenges that result in their failure or stagnation which causes an increase in the unemployment rate, poverty and inequality. A 2018 report by the Small Enterprises Development Agency (SEDA), an agency of the Department of Small Business Development with a mandate to develop small businesses, shows a year-on-year decline in the number of SMMEs in South Africa (Small Enterprises Development Agency, 2018). There are many contributing factors to the failure of small businesses, including a lack of information and communication technology (ICT) infrastructure, a lack of access to credit or financial support and inadequate managerial and employee skills and knowledge, among other factors.

In South Africa, the definition of an SMME is provided in the National Small Business Act (Act 102 of 1996), which defines a small business as "a separate and distinct entity, including cooperative enterprises and non-governmental organisations … which can be classified as a micro, a very small, a small or a medium enterprise" (The DTI, 2003). The National Small Business Act classifies an SMME by means of a matrix of business sector on the y-axis and size of full-time paid employees, total annual turnover and total gross asset value on the x-axis. In the classification of a small business, the number of employees is the same across all sectors: less than 5 for micro enterprises, less than 10 for very small enterprises and less than 50 employees and less than 200 for medium-size enterprises (except for the Agriculture sector where a medium-size enterprise has less than 100 employees). The values for total industry turnover differ according to sector. Stats SA, South Africa's national statistics agency, adjusts the total annual turnover cut off points for inflation to classify enterprises (Small Enterprises Development Agency, 2018). Table 1 shows the cut-off points – using the Rand currency – for enterprise turnover, as of December 2018.

SEDA (2018) segments SMMEs into either formal or informal. Enterprises operating in the formal sector are registered with the Companies and Intellectual Property Commission (CIPC), South Africa's companies' registration agency, and submit their annual tax returns to the South African Revenue Services (SARS); and SMMEs operating in the informal sector are not registered with the CIPC and comprise largely a single worker who is the owner. According to SEDA's SMME Quarterly Update report (2018) for the 1st quarter of 2018, out of the 2 443 163 SMMEs in South Africa, 73% are in the informal sector and 27% are in the formal sector.

The SMME sector is critical for the development of South Africa's economy: alleviating poverty, unemployment and inequality. The government's National Development Plan (NDP) recognizes that to achieve the objectives of the plan, among other initiatives, it needs to provide support for small businesses by easing rules for and procuring from small businesses, as well as improve access to finances for SMMEs (National Planning Commission, 2012). According to Stats SA's Quarterly Financial Statistics (QFS) for March 2018, SMMEs (excluding those with a turnover of less than R2 million) contributed 35% of South Africa's total turnover for the quarter (Statistics South Africa, 2018). And although the SMME Quarterly Update report (2018) indicates a 1.4% year-on-year decline in the number of SMMEs and 15.9% year-on-year decrease in the number of jobs they provide, this sector is still the larger provider of employment in the South African economy with the 8 886 015 jobs provided constituting around 55% of South Africa's total workforce.

There is a small number of studies about cloud computing adoption by SMMEs (Kumar *et al.*, 2017). In South Africa, there exists limited research about the opportunities and challenges of cloud computing adoption for SMMEs. Hinde & Belle (2012) recommended that a study with a more representative and larger sample be conducted in future to enrich the knowledge of the benefits and challenges of cloud computing for South African SMMEs. Kumalo & Poll (2015) recommend further research on cloud computing to address the challenges experienced by SMMEs. This research uses the TOE (technology-organisation-environment) framework to determine the challenges and opportunities of cloud computing for SMMEs in South Africa. The seeks to find the factors that represent opportunities and challenges of cloud computing adoption by the SMMEs. Formal SMMEs were selected for participation in the study as they are more likely to know about and use cloud computing and, due to the data collection method chosen, are more likely to be reachable for participation. The goal of this paper is to provide leaders of SMMEs an empirical study and resulting information about factors to take into consideration when making the decision whether adopt cloud computing.

| Industry Turnover | Medium (in millions) | Small (in millions) | Very small (in millions) |
|---|---|---|---|
| Mining and Quarrying | R 125 | R 50 | R 2 |
| Manufacturing | R163 | R 63 | R 2 |
| Electricity, Gas and Water | R163 | R 64 | R 2 |
| Construction | R 75 | R 38 | R 2 |
| Retail and Motor Trade Services | R 238 | R 50 | R 2 |
| Wholesale Trade, Commercial Agents and Allied Services | R 400 | R 50 | R 2 |
| Catering, Accommodation and other Trade | R 75 | R 64 | R 2 |
| Transport, Storage and Communication | R 163 | R 38 | R 2 |
| Real Estate and Business Services | R 163 | R 38 | R 2 |
| Community, Social and Personal Services | R 75 | R 13 | R 2 |

**Table 1: Stats SA: Cuff-off points for enterprise turnover to determine their size (December 2018)**

## 2. RELATED LITERATURE

This section discusses the definition of cloud computing, followed by the benefits and the opportunities presented by cloud computing, then discusses the Technology, Organisation and Environment (TOE) framework.

### 2.1 Definition of cloud computing

The concept of cloud computing has been in existence for some time. It started in the 1950s when dumb terminals connected to and accessed a complex and expensive mainframe computer over a network (Neto, 2014). The idea of cloud computing can be traced back to 1961 when Professor John McCarthy, an American computer scientist at Stanford University who created the computer time-sharing theory (McCarthy, 1983), said that "computing may someday be organised as a public utility …" (Garfinkel, 2011). In 1966, DF Parkhill further explored the forms and models – virtualization and grid computing – and benefits – latency of IT complexity and reduction in IT costs – of utility computing (Parkhill, 1966). Between the 1970's and the 1980's, cloud computing as a concept was advanced by the creation of the World Wide Web and the emergence of virtual machines (VMs) and virtual private networks (VPNs) by telecommunications companies (Neto, 2014). Salesforce is credited to be one of the first movers into cloud computing by delivering enterprise applications over the Internet in 1999 (Tripathi & Jigeesh, 2013).

### 2.2 Benefits of cloud computing for SMMEs

Enterprises must adapt to the constantly changing business environment by using cutting-edge technology that provide competitive advantage. Following are some of the main benefits that SMMEs can derive from cloud computing.

*Ease of management of IT systems.* Depending on the service model – IaaS, PaaS or SaaS – cloud computing reduces or eliminates the need for SMMEs to manage information technology systems. On a SaaS subscription, the cloud customer is not concerned with how or where the software is run and managed as those are hidden from them; the cloud service provider ensures that the software is always available to the customer. Thus, companies do not rely on internal infrastructure and footprint with regards to facilities (Vargas *et al.*, 2017).

*Reduction of IT costs.* Cloud computing is cost-effective for SMMEs (Adane, 2018). Affordability is the most attractive aspect of cloud computing for SMMEs, particularly in developing economies, as found by a survey conducted by Rath *et al.* (2012).

*Centralized working environment.* Cloud computing provides a centralized platform for development, testing, deployment, data storage and IT management for decentralized teams (Bartoletti, 2017). This improves team collaboration and resource sharing and removes duplication through integrated product teams that focus on creating customer-centric products. Korongo et al. (2013) asserted that cloud computing enables collaboration between cross-functional teams through resource sharing and increased usage of hardware, and that businesses can be conducted from any geographical location and at any time.

*Improve application delivery.* Cloud computing provides access to advanced middleware and data services that improve application delivery by increasing developer productivity and code quality through the reduction of errors, testing costs and an increase in accuracy (Vargas *et al.*, 2017).

*Scalability.* The scalability of computing resources is an important aspect of the cloud to businesses (Tripathi & Jigeesh, 2013). Cloud computing allows an SMME to scale up and down according to the computing requirements of the business and offers great opportunities to enable growth.

*Improved security.* Cloud computing removes the need for physical security required to protect IT infrastructure (Lalev, 2017). AWS (2017), in asserting that it has comprehensive security capabilities, explains six advantages of cloud security: 1) integration of compliance and security; 2) economies of scale where all AWS customers benefit from security innovation and improvements; 3) customers don't have to be concerned about managing security; 4) system configurations can be infused with all security features; 5) information about security issues are included; and 6) and using the cloud to protect the cloud service. Nedelcu *et al.* (2015) argue that although cloud computing may not be ideal for storing highly sensitive data, it is safer than on-premises systems.

*Improved performance and high availability.* Cloud computing enables easy and convenient anytime, anywhere access to data and applications using any type of device that has an internet connection (Khan, 2014), and therefore is always available.

*Competitiveness.* Sheedy (2018) found that enterprises are using the public cloud to be more competitive and rapidly create new customer value by leveraging the ecosystem of partners and independent service providers (ISVs) in cloud computing.

## 2.3 Theoretical background for cloud computing adoption

There are several theories used in studying the determinants of the adoption of technology by organisations: Technology Readiness (TR), Task-technology Fit (TTF), Technology Acceptance Model (TAM), TAM2, TAM3, Theory of Planned Behaviour (TBP), Decomposed Theory of Planned Behaviour, the Unified Theory of Acceptance and Use Technology (UTAUT), and Diffusion of Innovation (DoI) and Technology-Organisation-Environment (TOE) framework (Lai, 2017; Tarhini *et al.*, 2015). Lai reviewed the different technology adoption theories and concluded that different models and theories can be used based on the research problem, variables and measurement.

The DOI and the TOE models are the most popular theories is studying the factors that influence the adoption of technology. Diffusion of Innovation (DOI) was developed by Everitt Rogers (2003) to help the acceleration of the adoption and diffusion of new technological ideas through communication over time to remove uncertainty about the innovation. Cloud computing has been diffused and has been growing since the mid-2000s (Fabel, 2018). Low *et al.* (2011) found that factors that affect the adoption of cloud computing can be explored using the TOE framework, which, although different industries may show different determinant, covers all aspects of adoptions issues.

## 2.4 The TOE Framework

The TOE framework has been used in many studies to determine the factors that influence the adoption of cloud computing. Low *et al.*, (2011) used the framework to examine eight factors that determine the adoption of cloud computing: relative advantage, complexity, compatibility, top management support, firm size, technology readiness, competitive pressure, and trading partner pressure. Alshamaila *et al.* (2013) used TOE to study the cloud adoption process for SMEs in the north east of England. Li *et al.* (2015) explained the drivers of cloud service transformation in SMEs using the framework. Gangwar *et al.* (2015) integrated the TOE framework with the TAM (technology acceptance model) to understand cloud computing adoption. And many more other studies have been conducted using the TOE framework (Alshamaila *et al.,* 2013).

## 3. RESEARCH MODEL AND HYPOTHESIS DEVELOPMENT

This section proposes a research model based on the TOE framework and then presents the developed hypotheses. The model includes the technological, organisational and environmental contexts for factors that pose challenges and present opportunities of cloud computing for SMMEs. Figure 1 depicts the model, derived from Low et al. (2011), with the TOE drivers that determine the adoption of cloud computing.

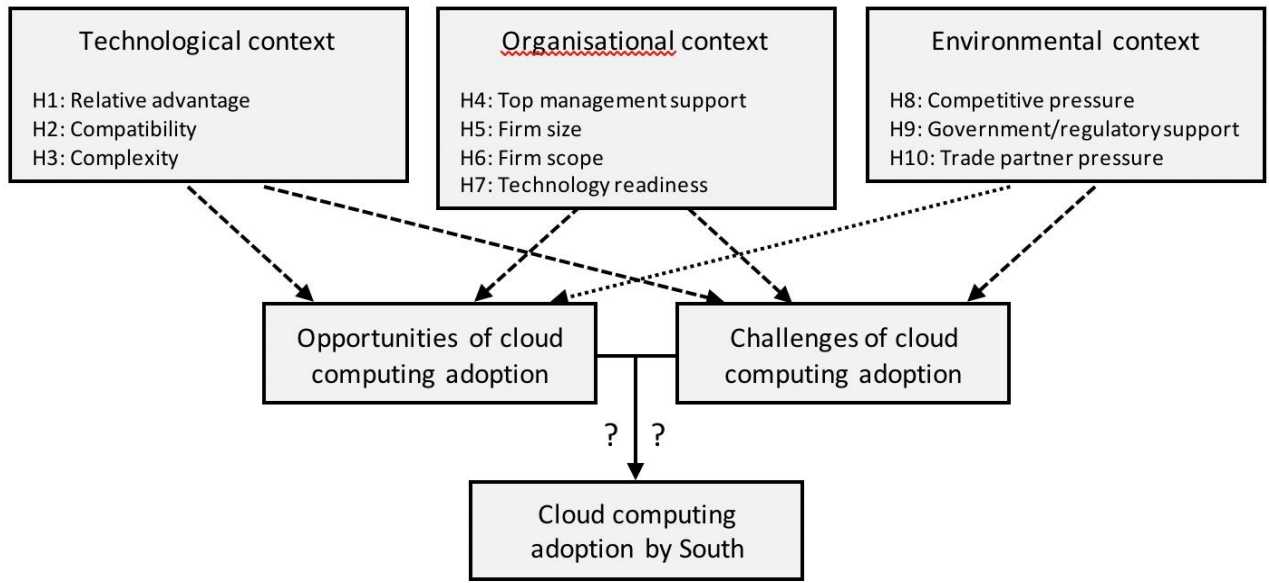

Figure 1: Research model

### 3.1 Technological context

The technological context includes both internal and external technologies that are relevant to the firm (Louis G. Tornatzky, 1990) and influence its decision to adopt a new technology. These refer to technology systems that are already being used by the enterprise – which determine the scope of its capabilities – and those that are available on the market and are relevant to the enterprise to improve its capabilities.

*Relative advantage* hypothesis

> *H1: There is a relationship between relative advantage and opportunities and challenges of cloud computing adoption by SMMEs in South Africa.*

*Compatibility* hypothesis

> *H2: There is a relationship between compatibility and opportunities and challenges of cloud computing adoption by SMMEs in South Africa.*

*Complexity* hypothesis:

> *H3: There is a relationship between complexity and opportunities and challenges of cloud computing adoption by SMMEs in South Africa.*

### 3.2 Organisational context

The organisational context refers to the characteristics – the firm's size, degree of centralization, degree of formalization, managerial structure – and resources – human resources, number of slack resources, and linkages among employees – of the firm (Louis G. Tornatzky, 1990).

*Top management support* hypothesis:

> *H4: There is an association between top management support and opportunities and challenges of cloud computing adoption by SMMEs in South Africa.*

*Firm size* hypothesis:

> *H5: There is an association between firm size and opportunities and challenges of cloud computing adoption by SMMEs in South Africa.*

*Sithole & Ruhode*                                               *Cloud Computing Adoption: Challenges and Opportunities for SMMEs**Firm scope* hypothesis:

> H6: There is an association between firm scope and opportunities and challenges of cloud computing adoption by SMMEs in South Africa

*Technology readiness* hypothesis:

> H7: There is an association between technology readiness and opportunities and challenges of cloud computing adoption by SMMEs in South Africa.

### 3.3    Environmental context

The environmental context refers to how the industry in which the firm does business is composed. Enterprises operate in a highly competitive environment where many other businesses offer similar and substitute products.

*Competitive pressure* hypothesis:

> H8: There is a relationship between competitive pressure and opportunities and challenges of cloud computing adoption by SMMEs in South Africa.

*Government and regulatory support* hypothesis:

> H9: There is a relationship between government and regulatory support and opportunities and challenges of cloud computing adoption by SMMEs in South Africa.

*Trading partner pressure* hypothesis:

> H10: There is a relationship between trading partner pressure and opportunities and challenges of cloud computing adoption by SMMEs in South Africa.

## 4.    DATA COLLECTION

Data was collected through an online survey questionnaire that comprised a formalized and prespecified list of close-ended questions with the option of entering free-text in some questions where the provided possible answers do not satisfy the survey participant. The questionnaire had two sections: company profile information and questions to evaluate the factors that affect cloud computing adoption. See appendix for the online survey questionnaire. The survey took was designed and tested to take approximately 10 minutes to complete.

To ensure that the questionnaire measures the objectives of the research and that consistent responses are obtained, the questionnaire was validated through face validity and content validity. Face validity will be performed by having two professional cloud computing subject matter experts who have before conducted research in any field read through the questionnaire to assess whether it effectively measures the topic it is intended to investigate. This will help eliminate poor-quality questions or add more relevant high-quality questions. Content validity was used determine whether the content of the survey is consistent, and effectively and expansively measures what the research purports to measure.

## 5.    DATA ANALYSIS AND RESULTS

The purpose of this study is to identify how cloud computing adoption can benefit small, medium and micro-sized enterprises (SMMEs) in South Africa. A quantitative research approach was adopted for this study to investigate the objectives.

### 5.1.    Hypotheses of the study

Based on the conceptual model discussed in the literature review, there were ten hypothesis that were evaluated for their relationship.

Proceedings of the 1st Virtual Conference on Implications of Information and Digital Technologies for Development, 2021

287

### 5.1.1. Nominal variables hypotheses

There was a total of three nominal variable hypotheses that were tested, and these variables were Q9_5 (Hypothesis 4), Q5 (Hypothesis 5) and Q8 (Hypothesis 6). The relationship for the nominal variables was investigated using Chi-square, with the strength with Cramer's V (φ) and the contribution to the significance with adjusted residual (Adj. residual > ±1.96).

Relationship between top management support and opportunities and challenges of cloud computing adoption

*H4a = There is an association between top management support and opportunities of cloud computing adoption*

The analysis was conducted to describe the relationship top management support using lack of access to finance as variable and opportunities of cloud computing adoption. The chi-square test was low with a value of 1.728, results show no significant relationship between the two variable, $\chi^2 = 1.728$ and a p-value of 0.189.

| Opportunities (Binned) | | | Disagree | Neutral | Agree | Strongly agree | Total | Chi-square | Sig |
|---|---|---|---|---|---|---|---|---|---|
| Lack of access to finance | Strongly Agree | Count | 3 | 5 | 11 | 22 | 41 | 1.728 | 0.189 |
| | | Adjusted Residual | 0.3 | 0.4 | 1.1 | -1.3 | | | |
| | Agree | Count | 2 | 10 | 13 | 41 | 66 | | |
| | | Adjusted Residual | -1.5 | 1.5 | -0.2 | 0.0 | | | |
| | Neutral | Count | 3 | 2 | 6 | 15 | 26 | | |
| | | Adjusted Residual | 1.1 | -0.5 | 0.3 | -0.5 | | | |
| | Disagree | Count | 2 | 0 | 4 | 18 | 24 | | |
| | | Adjusted Residual | 0.4 | -1.8 | -0.5 | 1.4 | | | |
| | Strongly Disagree | Count | 1 | 1 | 1 | 10 | 13 | | |
| | | Adjusted Residual | 0.2 | -0.4 | -1.2 | 1.1 | | | |

**Table 2: Relationship between top management support and opportunities to adopt cloud computing**

*H4b = There is an association between top management support and challenges of cloud computing adoption*

The association between top management support as measured by the lack of access to finance challenges and is presented on Table 3. The results show that there was no statistically significant association between challenges and lack of access to finance ($\chi^2 = 3.347$, p=0.067.

| ChallengesY2 (Binned) | | | Strongly disagree | Disagree | Neural | Agree | Strongly agree | Total | Chi-square | Sig | φ |
|---|---|---|---|---|---|---|---|---|---|---|---|
| Lack of access to finance | Strongly Agree | Count | 0 | 5 | 4 | 7 | 25 | 41 | 3.347 | 0.067 | 0.149 |
| | | Adjusted Residual | -0.8 | 1.7 | 1.0 | 0.8 | -1.9 | | | | |
| | Agree | Count | 2 | 4 | 3 | 10 | 47 | 66 | | | |
| | | Adjusted Residual | 1.8 | -0.2 | -0.8 | 0.5 | -0.3 | | | | |

| | | | | | | | |
|---|---|---|---|---|---|---|---|
| Neutral | Count | 0 | 2 | 1 | 1 | 22 | 26 |
| | Adjusted Residual | -0.6 | 0.3 | -0.6 | -1.6 | 1.5 | |
| Disagree | Count | 0 | 0 | 1 | 4 | 19 | 24 |
| | Adjusted Residual | -0.6 | -1.4 | -0.5 | 0.5 | 0.8 | |
| Strongly Disagree | Count | 0 | 0 | 2 | 1 | 10 | 13 |
| | Adjusted Residual | -0.4 | -1.0 | 1.4 | -0.6 | 0.4 | |

**Table 3: Relationship between top management support and challenges to adopt cloud computing**

In summary, there is no relationship between the top management support and the opportunities nor the challenges to the adoption of cloud computing.

Relationship between Firm size and opportunities and challenges of cloud computing adoption

*H5a = There is an association between firm size and opportunities of cloud computing adoption*

Table 4 presents the results that were conducted to understand the association between full-time employee and opportunities. The results show that there is no statistically significant relationship between the two models ($\chi2 = 0.053$, p =0.818).

| | Opportunities (Binned) | | Disagree | Neutral | Agree | Strongly agree | Total | Chi-square | Sig |
|---|---|---|---|---|---|---|---|---|---|
| How many full-time people does your company employ? | Less than 5 (micro) | Count | 5 | 7 | 12 | 34 | 58 | 0.053 | 0.818 |
| | | Adjusted Residual | 0.8 | 0.5 | 0.0 | -0.7 | | | |
| | Between 5 and 9 (very small) | Count | 2 | 4 | 4 | 26 | 36 | | |
| | | Adjusted Residual | -0.3 | 0.1 | -1.6 | 1.4 | | | |
| | Between 10 and 49 (small) | Count | 1 | 5 | 14 | 36 | 56 | | |
| | | Adjusted Residual | -1.7 | -0.5 | 1.0 | 0.4 | | | |
| | Between 100 and 199 (medium) | Count | 1 | 2 | 3 | 8 | 14 | | |
| | | Adjusted Residual | 0.1 | 0.5 | 0.1 | -0.4 | | | |
| | 200 or more (large) | Count | 2 | 0 | 2 | 2 | 6 | | |
| | | Adjusted Residual | 2.7 | -0.9 | 0.8 | -1.5 | | | |

**Table 4: Relationship between and firm size and opportunities of cloud computing adoption**

*H5b = There is an association between firm size and challenges of cloud computing adoption*

The analysis was conducted to describe the relationship between number of full-time employees and challenges, a chi-square was used to test the relationship. The chi-square test was low with a value of 0.008, results show no significant relationship between the two models (p>0.05), with $\chi2= 0.008$ and a p-value pf 0.928.

| ChallengesY2 (Binned) | Strongly disagree | Disagree | Neural | Agree | Strongly agree | Total | Chi-square | Sig | φ |
|---|---|---|---|---|---|---|---|---|---|

| How many full-time people does your company employ? | Less than 5 (micro) | Count | 0 | 3 | 5 | 9 | 41 | 58 | 0.008 | 0.928 | 0.119 |
|---|---|---|---|---|---|---|---|---|---|---|---|
| | | Adjusted Residual | -1.0 | -0.5 | 0.8 | 0.5 | -0.3 | | | | |
| | Between 5 and 9 (very small) | Count | 0 | 3 | 2 | 5 | 26 | 36 | | | |
| | | Adjusted Residual | -0.7 | 0.5 | -0.3 | 0.1 | 0.0 | | | | |
| | Between 10 and 49 (small) | Count | 1 | 5 | 3 | 7 | 40 | 56 | | | |
| | | Adjusted Residual | 0.5 | 0.9 | -0.4 | -0.3 | -0.2 | | | | |
| | Between 100 and 199 (medium) | Count | 1 | 0 | 1 | 1 | 11 | 14 | | | |
| | | Adjusted Residual | 2.2 | -1.0 | 0.1 | -0.7 | 0.5 | | | | |
| | 200 or more (large) | Count | 0 | 0 | 0 | 1 | 5 | 6 | | | |
| | | Adjusted Residual | -0.3 | -0.7 | -0.7 | 0.2 | 0.6 | | | | |

**Table 5: Relationship between firm size and challenges of cloud computing adoption**

In summary, there is no relationship between the firm size and the opportunities nor the challenges to the adoption of cloud computing.

Relationship between firm scope and opportunities and challenges of cloud computing adoption

*H6a = There is an association between firm scope and opportunities of cloud computing adoption*

Table 6 presents the results of relationship between the firm scope using the footprint of the clientele (location) and the opportunities; the results show a significance ($p<0.05$) for this model with $\chi^2 = 7.588$ and a p-value of 0.006. The strength of the association was measured with Cramer's V ($\varphi$) which was found to be 0.193, which indicate a weak association. Adjusted residual was conducted to understand the groups that contributes towards the significant association and the results show that the major contributor are organisations with international footprint and opportunities for adoption of cloud computing.

| | Opportunities (Binned) | | Disagree | Neutral | Agree | Strongly agree | Total | Chi-square | Sig | φ |
|---|---|---|---|---|---|---|---|---|---|---|
| Where is your company's clientele located? | Local (suburb, town or city) | Count | 0 | 0 | 5 | 13 | 18 | 7.588 | 0.006 | 0.193 |
| | | Adjusted Residual | -1.2 | -1.5 | 0.8 | 0.9 | | | | |
| | Regional (province) | Count | 1 | 3 | 3 | 12 | 19 | | | |
| | | Adjusted Residual | -0.2 | 0.8 | -0.5 | 0.1 | | | | |
| | National (South Africa) | Count | 3 | 6 | 18 | 50 | 77 | | | |
| | | Adjusted Residual | -1.2 | -1.1 | 0.8 | 0.6 | | | | |
| | International (Africa only) | Count | 0 | 3 | 4 | 11 | 18 | | | |
| | | Adjusted Residual | -1.2 | 0.9 | 0.2 | -0.1 | | | | |
| | International (Global) | Count | 7 | 6 | 5 | 20 | 38 | | | |
| | | Adjusted Residual | 3.4 | 1.2 | -1.3 | -1.4 | | | | |

**Table 6: Relationship between firm scope and opportunities to adopt cloud computing**

*H6b = There is an association between firm scope and challenges of cloud computing adoption*

The association between challenges and company's clientele location is presented on Table 7. The results show that there was no statistically significant association between challenges and company's clientele location (χ2 = 3.101, p=0.078).

| ChallengesY2 (Binned) | | | Strongly disagree | Disagree | Neural | Agree | Strongly agree | Total | Chi-square | Sig | φ |
|---|---|---|---|---|---|---|---|---|---|---|---|
| Where is your company's clientele located? | Local (suburb, town or city) | Count | 1 | 0 | 1 | 3 | 13 | 18 | 3.101 | 0.078 | 0.146 |
| | | Adjusted Residual | 1.8 | -1.2 | -0.2 | 0.4 | 0.0 | | | | |
| | Regional (province) | Count | 0 | 3 | 3 | 3 | 10 | 19 | | | |
| | | Adjusted Residual | -0.5 | 1.8 | 1.8 | 0.3 | -2.0 | | | | |
| | National (South Africa) | Count | 1 | 5 | 5 | 12 | 54 | 77 | | | |
| | | Adjusted Residual | 0.1 | 0.0 | 0.0 | 0.7 | -0.6 | | | | |
| | International (Africa only) | Count | 0 | 1 | 1 | 2 | 14 | 18 | | | |
| | | Adjusted Residual | -0.5 | -0.2 | -0.2 | -0.3 | 0.5 | | | | |
| | International (Global) | Count | 0 | 2 | 1 | 3 | 32 | 38 | | | |
| | | Adjusted Residual | -0.8 | -0.3 | -1.1 | -1.2 | 1.9 | | | | |

**Table 7: Relationship between firm scope and challenges of cloud computing adoption**

In summary, there is no relationship between the firm scope and the opportunities nor the challenges to the adoption of cloud computing.

### 5.1.2. Continuous variables hypotheses

There were seven continuous variable which were formed in constructs and used to test the relationship with opportunities and the challenges to the adoption of cloud computing. The hypotheses were analysed using Pearson correlation where they are normally distributed or Kendall's tau b where they are non-normal.

*5.1.2.1.Reliability and Normality of constructs*

Table 8 presents the reliability and the normality of the constructs. Based on the guidelines of George and Mallery, the reliability of compatibility and trading partner pressure were poor, and these were not used further in the study, as such hypothesis 2 (compatibility) and hypothesis 10 (trading partner pressure) could not be tested. The other constructs showed acceptable to excellent reliability as they were all higher than 0.7. In addition, these constructs were also normally distributed as their Skewness and Kurtosis were within the guidelines of ±2 by Hair et al (2010).

| Constructs | Cronbach's Alpha | N of Items | Reliability | Skewness | Kurtosis |
|---|---|---|---|---|---|
| Relative advantage | 0,912 | 11 | Excellent | 0,235 | 0,381 |
| Compatibility | 0,592 | 3 | Poor | | |
| Complexity | 0,809 | 5 | Good | 0,994 | 1,182 |
| Technology readiness | 0,750 | 2 | Acceptable | 1,297 | 0,512 |
| Competitive pressure | 0,880 | 2 | Good | 0,857 | 0,154 |
| Government and regulatory support | 0,754 | 3 | Acceptable | 1,553 | 1,827 |
| Trading partner pressure | 0,569 | 2 | Poor | | |
| Opportunities | 0.908 | 8 | Excellent | 0.367 | 0.303 |
| Challenges | 0.878 | 11 | Good | -0.075 | -1.054 |

**Table 8: Pearson correlation for opportunities of cloud computing adoption**

*5.1.2.2. Pearson correlation and linear regression*
*Opportunities of cloud computing adoption*

Table 9 presents the Pearson correlation for opportunities, relative advantage, complexity, technology readiness, competitive pressure and government and regulation support. The results show that the constructs have a significant relationship with opportunities p<0.05, except for government and regulation support which does not have a significant relationship with the opportunities (p>0.05). Relative advantage has a strong positive Pearson correlation with the opportunities r (170) =0.698, p<0.01 based on the guidelines of Pallant (2010). Furthermore, complexity had a medium positive and significant relationship with the opportunities r (170) = 0.308, p<0.01, there was also a strong positive significance between technology readiness and opportunities, r (170) = 0.561, p<0.01.

| | | Opportunities | Relative Advantage | Complexity | Technology Readiness | Competitive Pressure | Gov Regulatory Support |
|---|---|---|---|---|---|---|---|
| Opportunities | Pearson Correlation | 1 | | | | | |
| | Sig. (2-tailed) | | | | | | |
| | N | 170 | | | | | |
| Relative Advantage | Pearson Correlation | .698** | 1 | | | | |
| | Sig. (2-tailed) | 0,000 | | | | | |
| | N | 170 | 170 | | | | |
| Complexity | Pearson Correlation | .308** | .396** | 1 | | | |
| | Sig. (2-tailed) | 0,000 | 0,000 | | | | |
| | N | 170 | 170 | 170 | | | |
| Technology Readiness | Pearson Correlation | .212** | 0,145 | 0,022 | 1 | | |
| | Sig. (2-tailed) | 0,005 | 0,059 | 0,772 | | | |
| | N | 170 | 170 | 170 | 172 | | |
| Competitive Pressure | Pearson Correlation | .561** | .526** | 0,102 | .188* | 1 | |
| | Sig. (2-tailed) | 0,000 | 0,000 | 0,185 | 0,014 | | |
| | N | 170 | 170 | 170 | 170 | 170 | |
| | Pearson Correlation | 0,112 | 0,093 | 0,066 | .644** | .183* | 1 |
| Gov Regulatory Support | Sig. (2-tailed) | 0,145 | 0,228 | 0,391 | 0,000 | 0,017 | |
| | N | 170 | 170 | 170 | 172 | 170 | 172 |

\*\*. Correlation is significant at the 0.01 level (2-tailed).

\*. Correlation is significant at the 0.05 level (2-tailed).

**Table 9: Pearson correlation for opportunities of cloud computing adoption**

Table 10 presents the linear regression analysis for opportunities and relative advantage, complexity, technology readiness, competitive pressure and government and regulation support. The model summary significant and stable with F = 50.39, p<.001 with R-square of 0.550 and the adjusted R-square of 0.539 with a standard error of the estimate of 1.00917. This implies that the significant variables which are relative advantage (β = 0.510, p < .001) and competitive pressure (β =0.265, p < .001) explained 55.0% of the variance for the organisation for opportunities for adoption of cloud computing. This model is stable with no autocorrelation as the Durbin-Watson was 1.912 and no multicollinearity (VIF<10).

**Model Summary[b]**

| Model | R | R Square | Adjusted R Square | Std. Error of the Estimate | Durbin-Watson |
|---|---|---|---|---|---|
| 1 | .742[a] | .550 | .539 | 1.00990 | 1.895 |

a. Predictors: (Constant), Competitive Pressure, Complexity, Technology Readiness, Relative Advantage

b. Dependent Variable: Opportunities

**ANOVA[a]**

| Model | | Sum of Squares | df | Mean Square | F | Sig. |
|---|---|---|---|---|---|---|
| 1 | Regression | 205.578 | 4 | 51.394 | 50.392 | .000[b] |
| | Residual | 168.283 | 165 | 1.020 | | |
| | Total | 373.861 | 169 | | | |

a. Dependent Variable: Opportunities

b. Predictors: (Constant), Competitive pressure, Complexity, Technology Readiness, Relative advantage

**Coefficients[a]**

| Model | | Unstandardized Coefficients | | Standardized Coefficients | t | Sig. | Collinearity Statistics | |
|---|---|---|---|---|---|---|---|---|
| | | B | Std. Error | Beta | | | Tolerance | VIF |
| 1 | (Constant) | -1.218 | .578 | | -2.109 | .036 | | |
| | Relative Advantage | .587 | .077 | .516 | 7.670 | .000 | .603 | 1.658 |
| | Complexity | .103 | .079 | .075 | 1.305 | .194 | .828 | 1.208 |

| | | | | | | | |
|---|---|---|---|---|---|---|---|
| Technology Readiness | .066 | .041 | .086 | 1.609 | .110 | .961 | 1.040 |
| Competitive Pressure | .300 | .071 | .265 | 4.248 | .000 | .698 | 1.432 |

a. Dependent Variable: Opportunities

**Table 10: Regression of continuous variables and opportunities for adoption of cloud computing**

## 5.2    Challenges of cloud computing adoption

The Pearson correlation between relative advantage, complexity, technology readiness, competitive pressure and government and regulation support with challenges is presented on Table 11. The results show that there was a negative significant correlation between challenges and complexity, r (170) =-0.339, p<0.01, while the relation between challenges and government and regulation support was weak but positively significant, r (170) = 0.169, p<0.05. Furthermore, the relationship between relative advantage, r (170) = -0.042, p>0.05, technology readiness r (170) = 0.119, p>0.05 and competitive pressure, r (170) = 0.063, p>0.05 with challenges were not significant.

| | | Challenges | Relative Advantage | Complexity | Technology Readiness | Competitive Pressure | Gov Regulatory Support |
|---|---|---|---|---|---|---|---|
| Challenges | Pearson Correlation | 1 | | | | | |
| | Sig. (2-tailed) | | | | | | |
| | N | 170 | | | | | |
| Relative Advantage | Pearson Correlation | -0,042 | 1 | | | | |
| | Sig. (2-tailed) | 0,582 | | | | | |
| | N | 170 | 170 | | | | |
| Complexity | Pearson Correlation | -.339** | .396** | 1 | | | |
| | Sig. (2-tailed) | 0,000 | 0,000 | | | | |
| | N | 170 | 170 | 170 | | | |
| Technology Readiness | Pearson Correlation | 0,119 | 0,145 | 0,022 | 1 | | |
| | Sig. (2-tailed) | 0,123 | 0,059 | 0,772 | | | |
| | N | 170 | 170 | 170 | 172 | | |
| Competitive Pressure | Pearson Correlation | 0,063 | .526** | 0,102 | .188* | 1 | |
| | Sig. (2-tailed) | 0,413 | 0,000 | 0,185 | 0,014 | | |
| | N | 170 | 170 | 170 | 170 | 170 | |
| Gov Regulatory Support | Pearson Correlation | .169* | 0,093 | 0,066 | .644** | .183* | 1 |
| | Sig. (2-tailed) | 0,028 | 0,228 | 0,391 | 0,000 | 0,017 | |
| | N | 170 | 170 | 170 | 172 | 170 | 172 |

\*\*. Correlation is significant at the 0.01 level (2-tailed).

\*. Correlation is significant at the 0.05 level (2-tailed).

**Table 11: Correlation between challenges and five dimensions**

Table 12 presents the linear regression analysis for complexity and government regulatory support as they had a significant relationship based on the results of Pearson correlation. The overall model was significant, F = 14.901, p < .001, though the variance explained was low with $R^2$ = 0.151 (15.1%). This low variance indicated that this model results must be treated with causing, even though there was no multicollinearity (VIF ~1.00) and Durbin Watson for autocorrelation was 2.093, showing not autocorrelation. Complexity (β =-0.351, p <0.01) contributed negatively to the variance explained while government regulatory support (β =0.192, p <.01) contributed positive to the challenges for the adoption of cloud computing.

**Model Summary[b]**

| Model | R | R Square | Adjusted R Square | Std. Error of the Estimate | Durbin-Watson |
|---|---|---|---|---|---|
| 1 | .389[a] | .151 | .141 | 2.43890 | 2.093 |

a. Predictors: (Constant), Gov Regulatory Support, Complexity

b. Dependent Variable: ChallengesY2

**ANOVA[a]**

| Model | | Sum of Squares | df | Mean Square | F | Sig. |
|---|---|---|---|---|---|---|
| 1 | Regression | 177.265 | 2 | 88.632 | 14.901 | .000[b] |
| | Residual | 993.354 | 167 | 5.948 | | |
| | Total | 1170.619 | 169 | | | |

a. Dependent Variable: ChallengesY2

b. Predictors: (Constant), Gov Regulatory Support, Complexity

**Coefficients[a]**

| Model | | Unstandardized Coefficients B | Std. Error | Standardized Coefficients Beta | t | Sig. | Collinearity Statistics Tolerance | VIF |
|---|---|---|---|---|---|---|---|---|
| 1 | (Constant) | 8.046 | .758 | | 10.616 | .000 | | |
| | Complexity | -.853 | .173 | -.351 | -4.920 | .000 | .996 | 1.004 |
| | Gov Regulatory Support | .304 | .113 | .192 | 2.686 | .008 | .996 | 1.004 |

a. Dependent Variable: ChallengesY2

**Table 12: Regression of continuous variables and challenges for adoption of cloud computing**

# 6. DISCUSSION

The study was conducted to test whether each variable in the hypotheses presented either an opportunity and/or challenge of cloud computing adoption for SMMEs in South Africa.

## 6.1 Technological context

The strong positive relationship between relative advantage and opportunities of cloud computing adoption is consistent with findings by *Low et al.*, (2011), Alshamaila *et al.* (2013), (Oliveira & Martins, 2011), Kumar *et al.* (2017). A study by Senyo et al. (2016) for cloud computing adoption in a developing country (like South Africa) also found the relationship to be significant, but, contrary to the finding of this study, with a negative relation. Consistently, there is no significance for the relationship between relative advantage and challenges of cloud computing adoption, which leads to the conclusion that relative advantage, which represents benefits and opportunities such cost reduction, competitiveness, scalability, etc. of cloud computing, has a strong significance in influencing cloud computing adoption for SMMEs in South Africa.

The positive significance of the relationship between complexity and opportunities and the negative significant of the relationship between complexity and challenges indicate that complexity is a factor in cloud computing adoption. The result of complexity having a negative significance on cloud computing adoption by SMMEs in South Africa is consistent with Louw *et al.* (2011) and conflicting with studies by Oliveira & Martins (2011) and Gutierrez, Boukrami, & Lumsden (2015) who found complexity to be a barrier. The possible reason the SMMEs that participated in this study do not link cloud computing with complexity is because cloud computing is deemed to be easy to use and it simplifies IT systems management and application development and delivery. The cloud compatibility variable was not tested on the results of the study as they were found to have poor reliability.

## 6.2 Organisational context

The study found top management support to be insignificant to either opportunities and challenges of cloud computing by SMMEs in South Africa. While Gutierrez *et al.* (2015) congruently found top management support to not be a driver of cloud computing adopiton, the finding of this study is inconsistent with other previous studies (Low *et al.*, 2011; Alshamaila *et al.*, 2013; Senyo *et al.*, 2016; AL-Shboul, 2018). This study viewed top management support from only a financial support perspective while knowledge and skill was assessed from the perspective of the enterprise compatibility. This could be reason for the finding being inconsistent with earlier studies.

Firm size was found to not have any significance on opportunities and challenges of cloud computing adoption, consistent with Senyo *et al.* (2016) and AL-Shboul (2018) and inconsistent with Low *et al.* (2011). Out of the 174 study participants, only 3.4% firms had more than 200 employees and 8% had between 100 and 199 employees. 88.6% of the participants are small, which could explain why firm size had no significance in the study. Firm scope showed a weak positive significance with the opportunities of cloud computing adoption, contrasting findings by Senyo *et al*. (2016). This indicates that the wider the geographical scope of an SMME, the more likely it is adopt cloud computing as it is seen as an opportunity to reach more customers, centralize IT systems and improve collaboration among other benefits.

The study also found technology readiness to have a weak positive significance with the opportunities of cloud computing adoption. This finding is consistent with Oliveira and Martins (2010), Zhu *et al.* (2003), Senyo *et al.* (2016), while being inconsistent with *Louw et al.* (2011) who unexpectedly found this factor to be insignificant. Technology readiness refers to the readiness of the people in the organisation and the required infrastructure to adopt cloud computing. This study shows that the SMMEs have the required capabilities to adopt cloud computing.

| Hypothesis | | Relationship | Finding | Navigation | Strength |
|---|---|---|---|---|---|
| H1 | H1a | Relationship between relative advantage and opportunities | Significant (p<0.05) | Positive | Strong |
| | H1b | Relationship between relative advantage and challenges | Not significant (p>0.05) | | |
| H3 | H3a | Relationship between complexity and opportunities | Significant (p<0.05) | Positive | Medium |
| | H3b | Relationship between complexity and challenges | Significant (p<0.05) | Negative | Weak |
| H4 | H4a | Relationship between top management support and opportunities | Not significant (p>0.05) | | |
| | H4b | Relationship between top management support and challenges | Not significant (p>0.05) | | |
| H5 | H5a | Relationship between firm size support and opportunities | Not significant (p>0.05) | | |
| | H5b | Relationship between firm size and challenges | Not significant (p>0.05) | | |
| H6 | H6a | Relationship between firm scope and opportunities | Significant (p<0.05) | Positive | Weak |
| | H6b | Relationship between firm scope and challenges | Not significant (p>0.05) | | |
| H7 | H7a | Relationship technology readiness and opportunities | Significant (p<0.05) | Positive | Weak |
| | H7a | Relationship between technology readiness and challenges | Not significant (p>0.05) | | |
| H8 | H8a | Relationship between competitive pressure and opportunities | Significant | Positive | Medium |
| | H8b | Relationship between competitive pressure and challenges | Not significant (p>0.05) | | |
| H9 | H9a | Relationship between government and regulatory support and opportunities | Not significant (p>0.05) | | |
| | H9b | Relationship between government and regulatory support and challenges | Significant (p<0.05) | Positive | Weak |

**Table 13: Results of the relationships**

## 6.3  Environmental context

The study found competitive pressure to have a positive significance on the opportunities of cloud computing adoption, consistent with studies by Louw *et al.* (2011) and Senyo *et al.* (2016). This finding could be attributable to the fact that SMMEs want to be competitive through the attraction of new customers and retention of existing customers by improving service levels and customer experience. Also, if SMMEs think or know that their competitors are adopting cloud computing in order to derive the advantages it offers, they are more likely to adopt the same technology. In the study competitive pressure was not seen by SMMEs as a challenge to cloud computing adoption.

Government and regulatory support had showed a weak positive significance on the challenges and no significance on the opportunities of cloud computing adoption by South African SMMEs, according to the study, meaning that government and regulatory support is perceived negatively. This is consistent with Zhu *et al*. (2003) and AL-Shbou (2018) who also found regulatory support to have significance. However, it is incongruent with Senyo *et al*. (2016) who found that this factor had no significant. In South Africa, this could be due to the fact that government is perceived to not be supportive to the success of SMMEs in regard to technology. Regulations, such as affirmative action and labour laws, are generally perceived to be barriers to the success of SMMEs because they make it hard to easily hire and fire, in the case of labour laws, and for SMMEs get certain work contracts, in the case of affirmative action.

# 7.    CONCLUSION

This study sought to investigate the factors that factors that represent opportunities and challenges of cloud computing adoption by the SMMEs. A literature review on cloud computing to provide a definition, benefit and opportunities and challenges of cloud computing and on the theoretical framework, the TOE framework, was performed to contextualize the factors that impact cloud computing adoption opportunities and challenges. The factors of the TOE framework was used to develop the research model and formulate 10 hypotheses that postulated the relationship between each of the factors and the opportunities and challenges of cloud computing adoption by SMMEs in South Africa: relationship between relative advantage and opportunities and challenges of cloud computing adoption by SMMEs in South Africa; relationship between complexity and opportunities and challenges of cloud computing adoption by SMMEs in South Africa; relationship between top management support and opportunities and challenges of cloud computing adoption by SMMEs in South Africa; relationship between firm size support and opportunities and challenges of cloud computing adoption by SMMEs in South Africa; relationship between firm scope and opportunities and challenges of cloud computing adoption by SMMEs in South Africa; relationship technology readiness and opportunities and challenges of cloud computing adoption by SMMEs in South Africa; relationship between competitive pressure and opportunities and challenges of cloud computing adoption by SMMEs in South Africa; and relationship between government and regulatory support and opportunities and challenges of cloud computing adoption by SMMEs in South Africa.

The results of the study found six factors to have significance on the opportunities and challenges of cloud computing adoption by SMMEs in South Africa: relative advantage, complexity, technology readiness, firm scope, competitive pressure and government support. Relative advantage, firm scope, technology readiness and competitive pressure were found to favourably represent opportunities of cloud computing adoption, while government support was found to present a challenge for cloud computing adoption. The complexity of cloud computing was significant as both an opportunity and a challenge; however, negatively as a challenge and positively as an opportunity, indicating that SMMEs did not find complexity as a problem for cloud computing adoption.  Two variables – top management support and firm size – were found to be in insignificant.  SMMEs consider relative advantage – the benefits and opportunities – to be a strong significant factor for cloud computing adoption.

# REFERENCES AND CITATIONS